\documentclass[sigconf]{acmart}

\usepackage{booktabs} 

\setcopyright{rightsretained}

\acmDOI{10.475/123_4}

\acmISBN{123-4567-24-567/08/06}

\acmConference{bigdas at KDD 2017}{Halifax, Nova Scotia}{Canada} 
\acmYear{2017}
\copyrightyear{2017}

\acmPrice{15.00}

\begin{document}
\title{Computational Motility Tracking of Calcium Dynamics in \textit{Toxoplasma gondii}}


\author{Mojtaba S. Fazli}
\orcid{0000-0002-6082-2538}
\affiliation{%
  \institution{Department of Computer Science, University of Georgia}
  \streetaddress{30602}
  \city{Athens} 
  \state{Georgia, 30602} 
  \postcode{30602}
}
\email{Mojtaba@uga.edu}

\author{Stephen A. Vella}
\affiliation{%
  \institution{Department of Microbiology, University of Georgia}
  \streetaddress{}
  \city{Athens} 
  \state{Georgia, 30605} 
  \postcode{30605}
}
\email{sav28290@uga.edu}

\author{Silvia N.J.Moreno}
\affiliation{%
  \institution{Department of Cellular Biology, University of Georgia}
  \streetaddress{}
  \city{Athens} 
  \state{Georgia, 30605}
  \country{}}
\email{smoreno@uga.edu}

\author{Shannon Quinn}
\affiliation{%
  \institution{Department of Computer Science, University of Georgia}
  \city{Athens}
  \state{Georgia, 30602}
  \country{}}
  \email{spq@uga.edu}

\begin{abstract}
\textit{Toxoplasma gondii} is the causative agent responsible for toxoplasmosis and serves as one of the most common parasites in the world. For a successful lytic cycle, \textit{T. gondii} must traverse biological barriers in order to invade host cells, and as such, motility is critical for its virulence. Calcium signaling, governed by fluctuations in cytosolic calcium (Ca\textsuperscript{2+}) concentrations, is utilized universally across life and regulates many cellular processes, including the stimulation of \textit{T. gondii} virulence factors, such as motility. Therefore, increases in cytosolic Ca\textsuperscript{2+}, called calcium oscillations, serve as a means to link and quantify the intracellular signaling processes that lead to \textit{T. gondii} motility. Here, we describe our work extracting, quantifying, and modeling motility patterns of \textit{T. gondii} before and after Ca\textsuperscript{2+} stimulation via the addition of pharmacological drugs and/or extracellular calcium. We demonstrate a computational pipeline including a robust tracking system using optical flow and dense trajectory features to extract \textit{T. gondii} motility patterns. Using this pipeline, we were able to track \textit{T.gondii} motility dynamics in response to cytosolic Ca\textsuperscript{2+} fluxes in extracellular parasites. This allows us to study how Ca\textsuperscript{2+} signaling via release from intracellular Ca\textsuperscript{2+} stores and/or from extracellular Ca\textsuperscript{2+} entry relates to motility patterns, a crucial first step in developing countermeasures for \textit{T. gondii} virulence.
\newline
\end{abstract}

%
%






\maketitle

\section{Introduction}

Cellular motility serves as a fundamental and core characteristic shared across life, with key molecular components conserved across protozoa, bacteria, and vertebrates. Within cellular biology directional motility is an essential process involved in embryonic development, wound healing, neurology, T-cell immune response, and tissue development \cite{pollard2003cellular}. \textit{Toxoplasma gondii} is the causative agent of disseminated toxoplasmosis. \textit{T. gondii} is capable of infecting virtually any nucleated cell, making it the most prolific of all parasitic infections. Infections are lifelong and the immunocompromised are at particularly high risk due to reactivation of dormant tissue cysts that can lead to complications such as encephalitis, myocarditis, ocular toxoplasmosis, and congenital transmission. Calcium signaling is utilized universally across
life, as binding of Ca\textsuperscript{2+} to signaling effectors induces a cascade of downstream processes. Fluxes in basal cytosolic Ca\textsuperscript{2+} levels ([Ca\textsuperscript{2+}]\textsubscript{i}) serve as the basis for signaling, as mechanisms to either rapidly increase or decrease cytosolic Ca\textsuperscript{2+} levels are intricately balanced. Calcium signaling has been shown to stimulate every step of the \textit{T. gondii} lytic cycle, comprised of egress, gliding motility, invasion, and replication. Nevertheless, little is known about the molecules and mechanism involved in Ca\textsuperscript{2+} signaling in \textit{T. gondii}. Egress, gliding motility, and invasion all require activation of the molecular machinery involved in generating the mechanical force needed for progression throughout the lytic cycle. This activation is achieved either directly or indirectly via secondary signaling messangers, orginating from upstream Ca\textsuperscript{2+} mediated processes. Unfortunately, the exact sequence and identity of the signaling molecules involved in \textit{T. gondii} Ca\textsuperscript{2+} signaling are unknown \cite{wang2015seroprevalence}\cite{borges2015calcium}.

To elucidate the connection between Ca\textsuperscript{2+} signaling and its effects on \textit{T. gondii} motility, we propose a novel motility tracking algorithm. Using this algorithm, we can track Ca\textsuperscript{2+} dynamics in extracellular parasites in order to determine how changes induced by Ca\textsuperscript{2+} signaling relate to motility patterns of invasive \textit{T. gondii}. Ideally, our work will help us identify differences between mammalian Ca\textsuperscript{2+} signaling and parasite Ca\textsuperscript{2+} signaling for future drug targets, shedding further light on the eventual development of countermeasures against \textit{T. gondii} virulence.

The impetus for this research was to design a robust tracking system capable of tracking and quantifying induced motility changes of \textit{T. gondii} parasites in response to Ca\textsuperscript{2+} signaling. Object tracking, the process of locating a moving object in video data, has been employed in a variety of applications, including human-computer interaction, security, surveillance, video communication, augmented reality, traffic control, medical imaging, and video editing \cite{mountney2010three}. Numerous examples of video tracking exist in biology and neuroscience literature, including an automated tracking and segmentation framework by Hue-Fang et al. \cite{yang2009cell}, a mathematical approach for cell tracking, through formulating the cell tracking problem which is proposed by Blazakis et al., \cite{blazakis2014optimal} and blood cell tracking using Hough transform model and fuzzy curve tracing \cite{kiratiratanapruk2012worm}.

Motility is a fundamental process across both unicellular and multicellular organisms, filling a critical role in processes such as cancer progression, embryonic development, immune response, and tissue development. Cellular homeostasis involves responding to numerous constantly changing environmental cues such as ionic composition or signaling compounds. Such stimuli induce the intracellular signaling pathways needed to activate the molecular machinery for generation of mechanical force. Thus, cellular motility is key to proper cellular homeostasis and development, and the molecules and mechanisms involved in the signaling events are of great interest \cite{becker2005world}. For these reasons, we examined the motility of \textit{T. gondii} and propose a robust tracking algorithm for the ultimate purpose of using eventual motility models to design therapeutic countermeasures.
\newline
Here, we proposed an algorithm to track and statistically analyze Ca\textsuperscript{2+} signaling-induced motility events and corresponding changes in motility patterns of \textit{T. gondii}. Toxoplasmosis is a disease caused by the \textit{T. gondii} parasite. Infections of toxoplasmosis are usually asymptomatic in healthy adults, and are characterized by occasional mild flu-like symptoms such as muscle aches and tender lymph nodes. In a small number of people, eye problems may develop. In those with a weak immune system severe symptoms such as seizures and poor coordination may occur. If infected during pregnancy, a condition known as congenital toxoplasmosis may affect the child and lead to still-born death \cite{wang2015seroprevalence}. These disease pathologies of \textit{T. gondii} motivate the need for an accurate and robust method for analyzing the source of \textit{T. gondii} virulence: its motility.

We propose a Ca\textsuperscript{2+} signaling-induced motility analysis model as described below. First, we developed a robust video tracker to identify and track \textit{T. gondii} parasites. Then we extracted the necessary statistics together with the spatial trajectories of the parasites. The purpose of the trajectories is to describe the motion path of the parasites.
\section{Method}
In this section, we discuss the methodology of our proposed computational pipeline.

\subsection{Data}
We recorded fluorescence microscopy videos of Ca\textsuperscript{2+} dynamics of \textit{T. gondii} expressing a genetically encoded calcium indicator (GCaMP6f).  In the presence of increased cytosolic Ca\textsuperscript{2+}, the indicator increases in fluorescence, allowing us to track Ca\textsuperscript{2+}
dynamics across the lytic cycle. Videos are acquired using a LSM 710 confocal microscope in a heated chamber set to $37^\circ$C. 35  mm  glass  bottom  cover dishes (MATTEK) were treated with 2 mL of 10\% FBS  (Fetal  Bovine  Serum)  in  PBS  (Phosphate Buffer  Solution)  pH  7.4  overnight  at  $4^\circ$C. The following  morning  excess  FBS  was  washed  with PBS. Freshly egressed parasites expressing GCaMP6f were collected, purified, and resuspended in 1 mL of Ringer Buffer without calcium. ~2.5 x 10\textsuperscript{7} parasites were loaded onto cover dishes using a cell culture cylinder and incubated for approximately 15 min on ice to allow for cells to adhere. After removing the cell culture cylinder, Ringer Buffer was added to a 2 ml final volume. [11]

\subsection{Software}
We implemented the pipeline in Python 2.7, including libraries such as Numpy, Scipy, and Matplotlib for optimized linear algebra computations and visualizations. The core of our tracking algorithm used a combination of tools available in the OpenCV 3.1 computer vision library.  The full code for our pipeline is openly available under the MIT open source license at https://github.com/quinngroup/toxoplasma

\subsection{Computational Pipeline}
Our pipeline is illustrated in Fig. 1. Broadly, we imported the videos, preprocessed them, and extracted the motility statistics, and parasite trajectories using two different trackers. With the manual tracking approach, we extracted the motility, intensity, and velocity of specific, manually-identified objects. In both cases, trajectories were extracted using optical flow analysis. 
After importing video data, we proceed with the following preprocessing steps:
\begin{description}
  \item[$\bullet$ Step 1:] {\textit{Apply gamma correction and logarithmic correction on one level of tracking to increase the chance of object recognition.}}
  \newline
  \item[$\bullet$ Step 2:] {\textit{Apply erosion and dilation after creating binary images to remove false negatives and false positives for each frame. Additionally, thinning the edges to distinguish any cells clumped together.}}
  \newline
  \item[$\bullet$ Step 3:] {\textit{Apply a localized histogram transformation to remove background hue.}}
\end{description}

In the next step, we designed two trackers to follow individual parasites and link them to Ca\textsuperscript{2+} oscillations. First, we started with a manual tracker: a baseline approach that followed the current state-of-the-art in \textit{T. gondii} Ca\textsuperscript{2+}-induced motility tracking. We then implemented a variant of the KLT tracker \cite{amini2014fast} to follow all the parasite objects in a given video simultaneously.  

\begin{figure}
\centering
\includegraphics[width=0.4\textwidth]{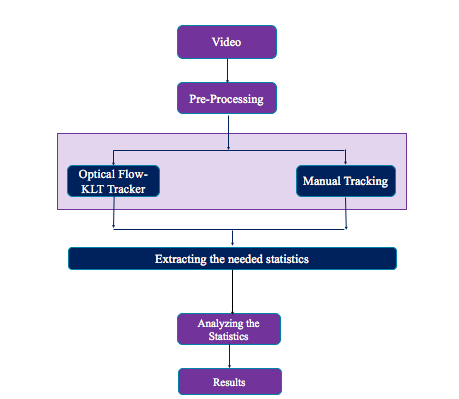}
\caption{\label{fig:Screen_Shot_2017-06-02_at_1_23_23_PM.png}\textit{Ca\textsuperscript{2+}-signaling induced motility tracking pipeline.}}
\end{figure}
\subsubsection{Manual Tracker}
The manual tracker requires tracking two versions of the data at once: grayscale and binarized levels. This is due to the fact that intensity serves as a direct correlation to cellular Ca\textsuperscript{2+} levels, in addition to the fluorescent \textit{T. gondii} parasites themselves. In our videos the cells expressing the genetically encoded Ca\textsuperscript{2+} indicator initially appear dim green, but after increases in cytosolic calcium via extracellular Ca\textsuperscript{2+} addition or addition of pharmacological drugs, causes the indicator to increase its fluorescence. Therefore, the intensity of each cell can be used to track calcium oscillations, and serves as the main factor we have to measure in relating Ca\textsuperscript{2+} signaling to \textit{T. gondii} virulence. Additionally, we track the motility patterns of the \textit{T. gondii} parasites themselves, in particular to observe the differences in motility before and after calcium stimulation. 

There are numerous disadvantages to a manual tracker in this context. First, the cells are abnormally shaped (banana-shaped) and undergo affine transformations throughout the video, making them difficult to follow using static shape templates. Second, they can move in and out of the focal plane of the microscope, causing the tracker to lose them. Third, minor oscillations in  fluorescence, irrespective of external Ca\textsuperscript{2+} addition or drug stimulation, causes the tracker to lose the objects. 
\newline
\begin{figure}
\centering
\includegraphics[width=0.3\textwidth]{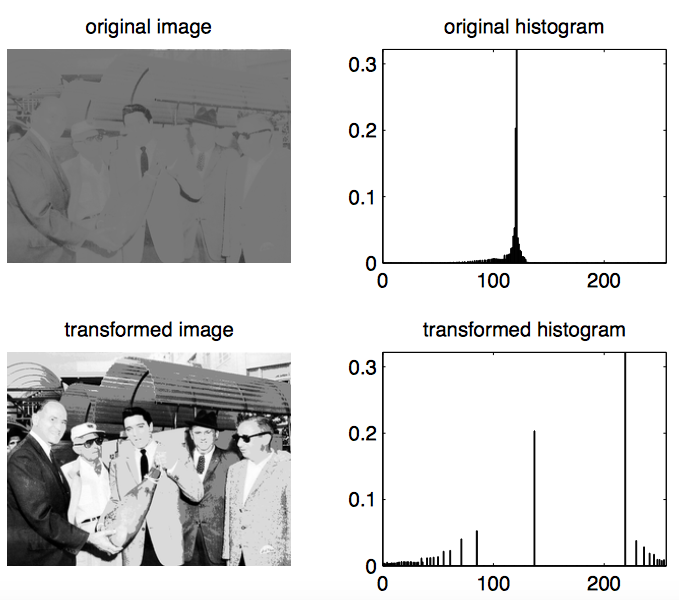}
\caption{\label{fig:Screen_Shot_2017-05-02_at_10_33_31_PM.png}\textit{Effect of localized histogram method applied on a sample image[6].}}

\end{figure}
To cope with these issues, we improved the algorithm in two steps. First, we applied some preprocessing operations like gamma correction and logarithmic correction. In this case, even if an object has a very low intensity we can recover it from the background. Then, we applied histogram localizations to remove background noise. In some cases, the background includes some green hue which may cause some noise. To eliminate this background noise, we used local histogram equalization.

Local histogram equalization\cite{ketcham1974image} is a technique for adjusting image intensities to enhance contrast. It differs from ordinary histogram equalization in the respect that the adaptive method computes several histograms, each corresponding to a distinct section of the image, and uses them to redistribute the lightness values of the image. It is, therefore, suitable for improving the local contrast and enhancing the definitions of edges in each region of an image \cite{ketcham1974image}. The application of this method is demonstrated in Fig. 2.

\begin{figure}
\centering
\includegraphics[width=0.35\textwidth]{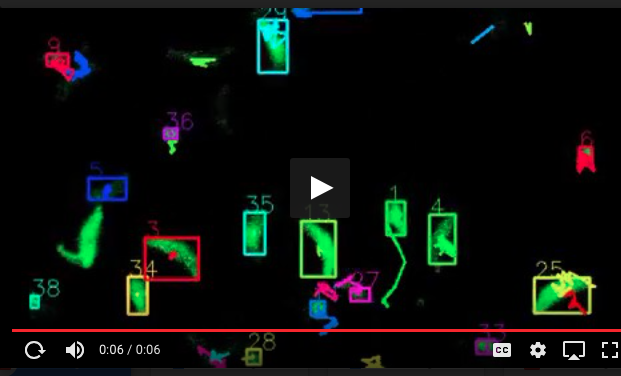}
\caption{\label{fig:Screen_Shot_2017-03-31_at_6_44_43_PM.png} \textit{A snapshot of final manual tracking of \textit{T. gondii} parasites.}}
\end{figure}
After applying the preprocessing steps to each frame, we find the contours of each parasite for each frame, assign a bounding box to each contour, and find the closest object spatially from the previous frame. These two objects are then considered the same object across subsequent frames. A representative image of our tracking module is illustrated in Fig. 3. We also establish parameters to handle objects that move out of frame or too which move too quickly to be tracked. For the latter, if the relative centers of the two aligned objects across frames are spatially separate beyond a certain threshold, they remain as two distinct objects to our tracker. For the former, when an object disappears, we record the position of disappeared object and wait a certain number of frames to see if it reappears. As with the object alignment, we place a hard threshold on the wait time; beyond that threshold we consider the object lost. 
After developing our manual tracking algorithm, we measured the velocity, average distance traveled, and intensity of each tracked parasite. By considering the cells separately we measured three kinds of motions and the intensities of each cell. For velocity computation, we considered the pixels which are passed in each subsequent frame. The details are illustrated in Fig. 4.

\begin{figure}
\centering
\includegraphics[width=0.45\textwidth]{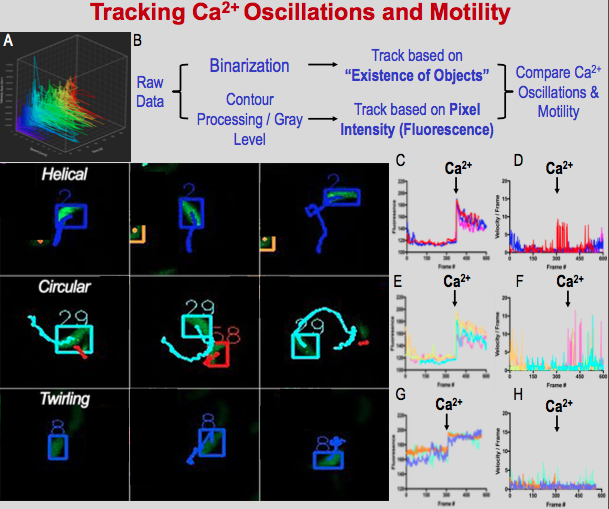}
\caption{\label{fig:Screen_Shot_2017-03-31_at_6_47_40_PM.png} \textit{Manual Tracking Result A) Calcium Oscillations vs. Time vs. Speed. B) Scheme for Tracking method. C) Representative images of motility types using new tracking module. C) Helical Ca\textsuperscript{2+} Oscillations Fluorescence Tracings D) Helical Velocity Tracks E) Circular Fluorescence Ca\textsuperscript{2+} Oscillations F) Circular Motility Velocity Tracks G) Twirling Ca\textsuperscript{2+} Oscillations Tracings H) Twirling Ca\textsuperscript{2+} Velocity Tracks} 
}
\end{figure}
\subsubsection{KLT Tracker}
Although our manual tracking algorithm is an improvement over that which was used in \textit{T. gondii} Ca\textsuperscript{2+}-induced motility assessment, we still needed a tracking approach robust enough to overcome some of the more common shortcomings of the manual tracker, such as parasite depth occlusions and alignment problems across frames. For this reason the KLT tracker was a good alternative to apply for our purposes. \cite{tomasi1991detection} The Lucas-Kanade tracker (KLT) uses the Tomasi algorithm to find fixed points. KLT makes use of the spatial intensity information to direct the search for the position that yields the best match. It is faster than traditional techniques for examining far fewer potential matches between images. In this approach, we consider some features extracted through the Tomasi algorithm of the object, which are fed to the KLT and tracked over the course of the video.\cite{aires2008optical} The resulting tracker is very robust against the kinds of failure modes we observed in our manual tracker (Fig. 5).   
\newline

\begin{figure}
\centering
\includegraphics[width=0.44\textwidth]{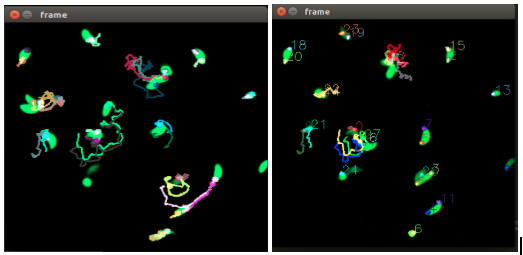}
\caption{\label{fig:Screen_Shot_2017-07-30_at_8_04_07_PM.png} \textit{Cell tracking result using KLT (left) and the same result by allocating numbers to each point (right) }}
\end{figure}

\section{Results}
We extracted trajectories from our manual tracking and the KLT implementation. Our findings are as follows. In Fig. 6 we observed all trajectories extracted from a sample video in a 2D space. Fig. 7 links this information with cytosolic Ca\textsuperscript{2+} oscillations, as indicated by the shade of the plot (brighter hue representing higher Ca\textsuperscript{2+} levels). From this, we are able to observe how the motility patterns of \textit{T. gondii} parasites abruptly change in the presence of extracellular Ca\textsuperscript{2+}, clearly implying a causal relationship.
\newline

\begin{figure}
\centering
\includegraphics[width=0.44\textwidth]{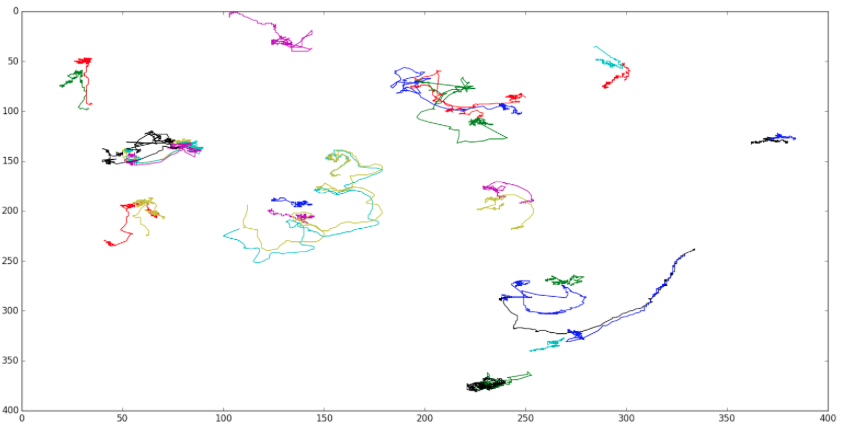}
\caption{\label{fig:Screen_Shot_2017-07-30_at_8_14_13_PM.png} \textit{2D plot of all cell trajectory using KLT (x and y axis are showing the coordinates of each point during the time)}}
\end{figure}

\begin{figure}
\centering
\includegraphics[width=0.44\textwidth]{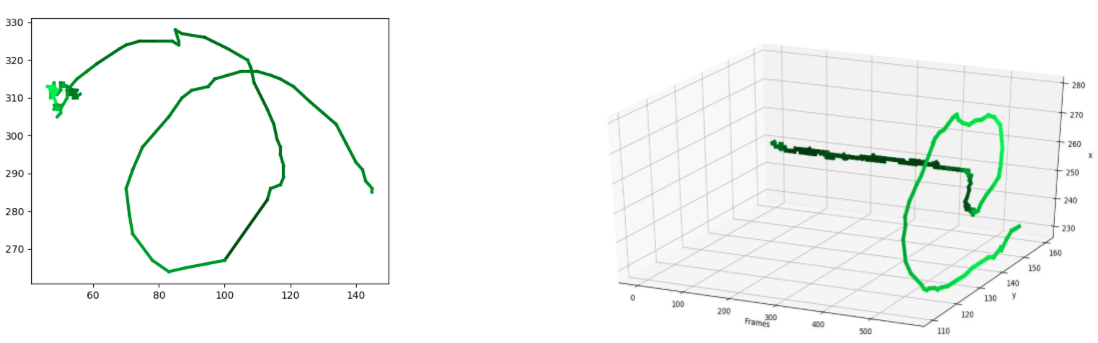}
\caption{\label{fig:Screen_Shot_2017-07-30_at_8_14_13_PM.png}\textit{2D plot of a sample cell trajectory with intensity and 3D representation of another sample object (left). On the right, the trajectory is colored to indicate the level of fluorescence of the tracked parasite. From this, it is easy to see when the Ca\textsuperscript{2+} signaling was added to the system.}}
\end{figure}

\subsection{Extracted statistics}
As a means to evaluate the robustness of our tracking algorithm against our hypothesis, we extracted the X and Y coordinates of a single motile cell.  This approach enables us to generate several pieces of statistical data related to motility, such as the translational and angular velocity, as indicated in Figures. 8 and 9.  Our cells exhibit multiple types of motility that relate to movement within a coordinate plane (helical and circular), thus it is necessary to compute both the translational and angular velocity, respectively. Furthermore, we can interpret how the fluorescence (cytosolic Ca\textsuperscript{2+} concentration), as indicated by the shade of the green hue of the line plot, relates to \textbf{burst} or \textbf{dips} in speed post addition of Ca\textsuperscript{2+} or drug stimulation. 

\subsubsection{Object Calcium Levels}
Levels of Ca\textsuperscript{2+} are indicated by the intensity of fluorescence of the tracked parasites. To compute the Ca\textsuperscript{2+} levels by way of intensity, for each point we considered a $nxn$ rectangle, while the coordination of the point is located at the center of that rectangle. Then we calculated the average intensity of all non-zero pixels located inside of the rectangle. The reason of considering a $nxn$ window instead of just one point intensity is clear: each point may be in the border of the object or even out of the object but attached to that object. Thus, by considering the intensity in a $nxn$ window we have better estimation of intensity value of an object.
\subsubsection{Object Velocities}

Object velocities were computed in two steps. First, we computed the instantaneous velocity of each object per pair of consecutive frames. To do this we needed to know the distance between 2 consecutive time points:
\newline
\begin{center}
$P1 = (x1, y1)$ \textbf{and} $P2 = (x2, y2)$ 	 \textbf{     (1)   }
\newline
\end{center}
Where $P1$ and $P2$ are the coordinates of an object across two consecutive frames. Thus, the distance between these two points is calculated through a simple Euclidean distance formula:

\[ d= \sqrt{(x_{1} - x_{2})^2 + (y_{1} - y_{2})^2} \textbf{     (2)   }\]
\newline
\[ v = \frac{d}{\Delta t} \textbf{     (3)   }\]
\newline

since we are going to check the velocity of each object per frame, we have $t = 1$. Thus, we can use the formula (2) as a metric to compute the instantaneous velocity. In the next step, we  smooth the instantaneous velocities with a sliding window. Our results in computing instantaneous and smoothed velocities are shown in Fig. 8.

\begin{figure}
\centering
\includegraphics[width=0.46\textwidth]{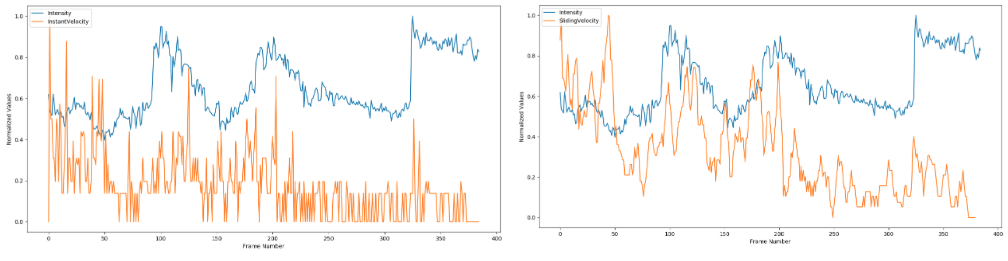}
\caption{\label{fig:Screen_Shot_2017-07-30_at_10_07_12_PM.png} \textit{2D plot of a sample cell intensity VS. Instantaneous velocity (left) and the averaged sliding velocity (right)}}
\end{figure}
\subsubsection{Objects Angular Velocities}
Angular velocity is an important component to understanding the instantaneous motility dynamics of \textit{T. gondii}, as our cells tend to move in counterclockwise circular patterns. To compute the angular velocity, we need to know the object orientation or object angle per frame, and we can calculate it as: 

\[
\theta = \arctan( \frac{y}{x})
\textbf{     (4)   }\]
\newline

where $y$ and $x$ are the coordinates of point $P$ in frame $T$. Thus after computing the angles of all points in all frames, we can easily compute the  Angular velocity which is defined as follows:
 \[
 \textit{Angular Velocity} = \theta_2 - \theta_1 \textbf{     (5)   }\]
\newline
Where $\theta_1$ and  $\theta_2$ are angle values of an object in $2$ consecutive frames. Thus, the differences between two angles in two consecutive frames shows instantaneous angular velocity. To obtain the sliding average angular velocity we should compute as done previously.

The significance of this metric is because of the fact that it can show the randomness of movements before calcium addition (if any random movement existed) and non-random motility after calcium addition. To compute the angle histogram, we should use the following formula: 

\[
\theta = \arctan( \frac{y_2 - y_1}{x_2 - x_1})
\textbf{     (6)   }\]
\newline

A representative plot of the instantaneous and sliding angular velocity using our tracking module is seen in Fig. 9. 
\begin{figure*}
\centering
\includegraphics[width=0.74\textwidth]{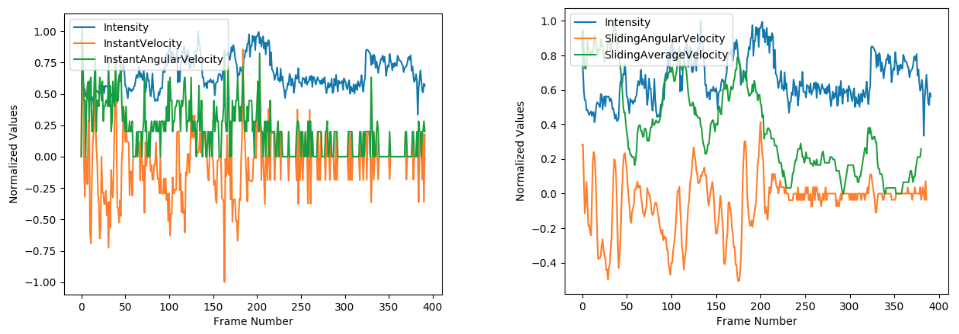}
\caption{\label{fig:Screen_Shot_2017-07-30_at_10_47_02_PM.png} \textit{ 2D plot of a sample cell intensity VS Instantaneous velocity and Instantaneous angular velocity (left) and the averaged sliding velocity and averaged sliding angular velocity(right)}}
\end{figure*}

\section{Discussion}

Trajectories play a crucial role in our work because we can follow the dynamics of cell motility across them. The pipeline presented here successfully identifies the three known motility types (Fig. 10). Part $A$ of Fig. 10 shows the motility types and the trajectories of three selected parasites found to be following each type. In part $B$,$C$, and $D$ of this figure one can see a $4D$ representations of each cell trajectory. These dimensions include $X$ and $Y$ coordinates, frame number, and the intensity of a representative cell track across consecutive frames as shown by the hue of the trajectory curve. The hue intensity represents the calcium levels of a single cell and is used as a proxy for $Ca^{2+}$ oscillations. At approximately the $350^{th}$ frame one can see an increase in the green hue as it becomes brighter, due to the addition of extracellular $Ca^{2+}$.  Afterwards, one see the cell begin to move.  It is important to highlight that increases in calcium (brighter hue coloring of the line) are linked to increases in motility ( a change in the $X$-$Y$ coordinates of the line).

From the XY coordinates we can extrapolate several pieces of information and relate them to the fluorescence signal (Ca\textsuperscript{2+} level).  If one looks at Fig. 10, part E indicates the calcium oscillation patterns for other similar objects undergoing the same type of motility as portrayed in their representative column. After the addition of Ca\textsuperscript{2+}, the fluorescence intensity is increased dramatically, as indicated by the addition of extracellular calcium around Frame 350 and appears to cause the cells to oscillate in a somewhat coordinated frequency and amplitude pattern. The most interesting finding is illustrated in Fig. 10 part H, where one can see the Ca\textsuperscript{2+} oscillations plotted together with the velocity across time, as peaks in normalized average velocity and normalized averaged angular velocity correlate with peaks of cytosolic Ca\textsuperscript{2+}. As it is clear in Fig. 10, most of the time, there is a direct relationship between changes in fluorescence intensity and the object translational and angular velocity. Moreover, if one looks at this part again he or she will see the addition of 2 kinds of reagents that are applied in 2 different points of the time - Thapsigargin (TG) (pharmacological reagent that causes increased levels of cytosolic Ca\textsuperscript{2+} and extracellular Ca\textsuperscript{2+})  - . What we discovered and found it extremely interesting was the effect of those two drugs on motility pattern; at the beginning the intensity is changing randomly, as a consequence the velocity and the angular velocity are also changing stochastically. Then by adding TG we will see the random oscillation of intensity is paused, causing the intensity, the object velocity, and the angular velocity to appear more stable - after addition of TG and before addition of Ca\textsuperscript{2+} -.  Finally by addition of Ca\textsuperscript{2+} , the intensity is increased sharply and after a short delay phase both the velocities are increased considerably. To highlight this phenotype in more detail we isolated and plotted the tracks within their own separate figure (Fig. 11). There on the left side of the figure, before addition of TG, the intensity is fluctuating randomly but by addition of TG around the 300th frame, then the intensity level becomes more stable. Again, we can see that by addition of Ca\textsuperscript{2+} at approximately the 400th frame, the intensity increases greatly. In the meantime, the translational velocity and angular velocity are shown on the right side of the figure. They show the same behavior as the intensity shows in front of TG and Ca\textsuperscript{2+} addition. Another interesting point is in the right side of the figure is at the frame 600, where, both the angular velocity and translational velocities increase dramatically a few seconds after reaching the max value of intensity caused by calcium stimulation.
\begin{figure}
\centering
\includegraphics[width=0.46\textwidth]{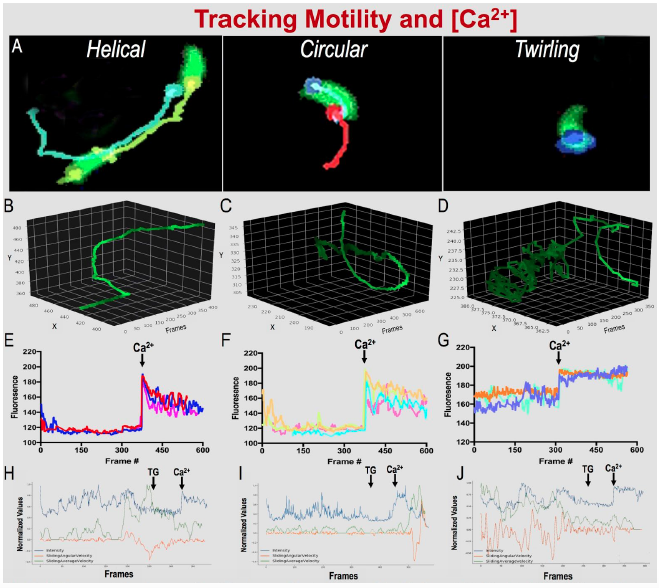}
\caption{\label{fig:Screen_Shot_2017-06-18_at_6_54_57_PM.png} \textit{A) Representative images of motility types using tracking module. B,C and D ) 4D plot of X and Y coordinates, Frame number, and fluorescence intensity ( line hue color ) of helical, circular, and Twirling motility, respectively. E,F, G) Ca\textsuperscript{2+} Oscillations Fluorescence Tracing of Helical, Circular and Twirling motility, respectively. H, I, and J ) Ca\textsuperscript{2+}Fluorescence Tracing versus Translation velocity versus Angular velocity of Helical, Circular, and Twirling motility respectively.
}}
\end{figure}

\begin{figure}
\centering
\includegraphics[width=0.45\textwidth]{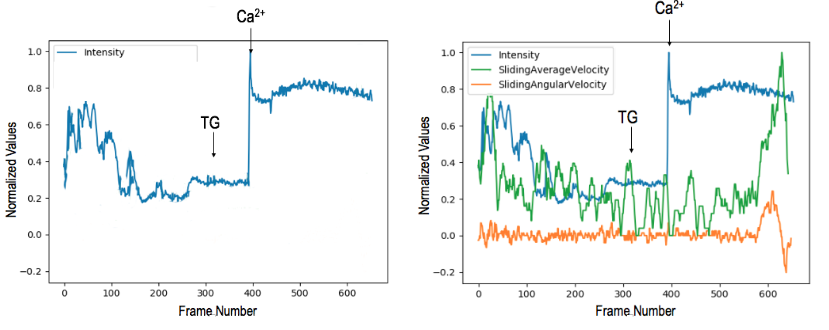}
\caption{\label{fig:Screen_Shot_2017-07-30_at_11_04_49_PM.png}\textit{left : The intensity-time plot of a specific cell , Right : the intensity vs translational velocity vs angular velocity. in both the sides the interesting points of TG and Ca\textsuperscript{2+} are indicated.
}}
\end{figure}

Fig. 12, shows those interesting points in a circular motility trajectory. This 4D representation from time view indicates The Ca\textsuperscript{2+} addition point and TG addition point.

\begin{figure}
\centering
\includegraphics[width=0.4\textwidth]{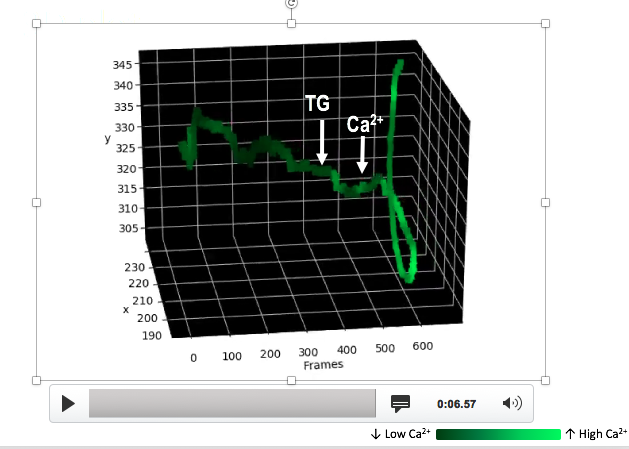}
\caption{\label{fig:Screen_Shot_2017-06-06_at_6_16_18_PM.png}\textit{left :  demonstration of TG and Ca\textsuperscript{2+} addition points on a 4D representation of a cell trajectory.
}}
\end{figure}
In addition, to see the interesting findings in an entire video, we depict all objects trajectories in a 4D view (Fig. 13) . This view, enables us to understand more of the effect of TG and Ca\textsuperscript{2+} . Moreover, As it is obvious at the same figure, after addition of calcium all the trajectories are much brighter than previous time which shows us the amount of calcium is increasing after calcium addition in all the cells.

\begin{figure}
\centering
\includegraphics[width=0.44\textwidth]{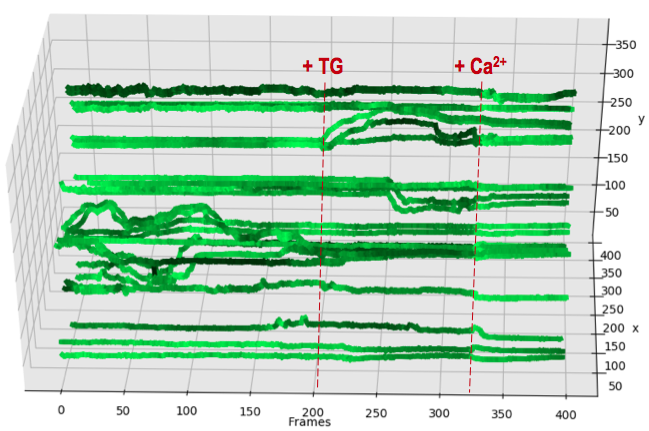}
\caption{\label{fig:Screen_Shot_2017-06-06_at_6_10_35_PM.png}\textit{Time-View demonstration of TG and Ca\textsuperscript{2+} addition points on a 4D representation of all cell trajectories.
}}
\end{figure}

\section{Generalizing the pipeline for big data and other types of trajectories}

In theory our framework's ability to track, analyze, and classify the motility dynamics of $Toxoplasma$ $gondii$ is readily applicable to related species of motile parasites, including $Plasmodium$ ookinetes and sporozoites that undergo similar motility patterns. However, our pipeline was designed from the ground-up with \textit{T. gondii} as it main applicant, and these are only initial results; there are numerous important applications of motility tracking throughout public health that would benefit from this framework, and we anticipate the application of this platform will be significantly useful to solve those problems. In order for our pipeline to be effectively applied to another organism, parameters such as the number frames recorded per second, the pixel size, the number of frames used for averaging speed, or determining angular velocity have to be carefully tested and optimized.

In our current application, the question is not simply a matter of \textbf{vast amount of video data}. Videos of \textit{T. gondii} motility are expensive to create and acquire. It requires fully-staffed labs with a lot of time to grow the parasites, plate them, and image them over periods of time. Hence, right now, the data-generation step is not scalable. Thus, we are not yet at the point of being able to say that the our video data are massive.  

However, in the future, we anticipate a scalable analysis framework in which  by rewriting the code in a distributed fashion (for instance, since our pipeline is written by python we can use easily PySpark or Flink) and managing the data we can generalize our algorithm to be robust on big data. To do that, first the videos would be analyzed in parallel to extract trajectories, then the trajectories would be stacked onto a tall-and-skinny matrix that is row-distributed. In this matrix,  each column and each row of a matrix correspond, respectively, to one object and one video frame. in each row, we measure the objects location ( X and Y value of objects ) and intensity value of the objects in that specific frame. So, here the objects are features and the rows are the observations of those features. Since the number of objects is quite fewer than the number of the frames we will have a tall and skinny matrix of trajectories. In the next step,  each trajectory could either be analyzed in parallel on its respective node, or we can perform distributed matrix decomposition on the matrix.

Generally speaking, our challenge with the big data would be rearrangement of distributed operations through Spark, Flink or Dask that can smoothly process each video chunk on HDFS and extract the desired statistics, exactly like what we have done previously on large scale FMRI images\cite{li2016scalable}.

\section{Conclusion}
Here, we developed a KLT-based tracker and compared it to manual tracking of Ca\textsuperscript{2+}-induced \textit{T. gondii} motility patterns. We shown that, how we extracted the data and statistics from trajectory. For future work, we already planned to capture the motions using dense trajectories. dense trajectories are the dense version of optical flow \cite{wang2011action} - despite the KLT which is the sparse version of optical flow - . So, We then study the trajectories by clustering them and applying machine learning techniques to categorize them. In doing so, we gain a deeper understanding of the possible motion types of \textit{T. gondii} under different environmental stimuli and, therefore, an avenue through which to understand the virulence factors of \textit{T. gondii}.

\bibliographystyle{ACM-Reference-Format}
\bibliography{sample-bibliography} 

\end{document}